\documentclass[twocolumn,showpacs]{revtex4} 

\usepackage{amsmath}
\usepackage{amssymb}
\usepackage{amsfonts}
\usepackage{graphicx}
\usepackage{bm}

\begin{document}

\title{Gibbs entropy of network ensembles by cavity methods}

\author{Kartik Anand$^1$ and  Ginestra Bianconi$^2$ }
\address{$^1$ Abdus Salam International Center for Theoretical Physics, Strada Costiera 11, 34014 Trieste, Italy\\
$^2$Department of Physics, Northeastern University, Boston 02115 MA, USA}

\begin{abstract} {
The Gibbs entropy of  a microcanonical network ensemble is the logarithm of the number of network configurations compatible with a set of hard constraints. This quantity characterizes the level of order and randomness encoded in features of a given real network. Here we show how to relate this entropy to large deviations of conjugated canonical ensembles. We derive exact expression for this correspondence using the cavity methods for the configuration model, for the ensembles with contraint degree sequence and community structure and for the ensemble with constraint degree sequence and number of lins at a given distance.
}
\end{abstract}

\maketitle

The evolution of complex networks are usually described by non-equilibrium stochastic dynamics \cite{Evolution1,Evolution2,Latora,Dorogovtsev,NewmanB}. However a networks' specific topological structure may reveal relevant organizational principles, such as an universality for the large-scale structure or hierarchical communities \cite{Santo} that is sure to impact dynamical processes taking place on the network \cite{Doro_c,Barrat}.

Extracting relevant statistical information encoded in the networks' structure is a fundamental concern of community detection algorithms \cite{Santo} and other inference problems.
To study these problems, several authors have suggested entropy based methods \cite{Gfeller,Features,Munoz}, which are grounded in the information theory of networks  \cite{entropy1, entropy2, BianconiC, Kartik, Annibale, Munoz}. These methods have proved to be very useful. In fact, in a series of recent papers \cite{entropy1, entropy2, BianconiC,Kartik,Annibale,Munoz,Newman1,Newman2,Newman3} it has been shown that one may extend ideas and concepts of statistical mechanics and information theory to complex network ensembles.

In this paradigm, one generalizes the typical random graph ensembles studied in the mathematical literature \cite{RG} to ensembles characterized by an extensive number of constraints that fix, for example, the degree sequence \cite{MR}, number of links between different communities, the number of links at a given distance \cite{entropy1,entropy2} ,degree correlations between linked nodes \cite{Munoz}, acyclic networks \cite{Newman1}, or even network with given number of triangles \cite{Newman2} and generalized motifs \cite{Newman3}. 
 
 It is well known that in statistical mechanics we distinguish between microcanonical ensembles describing all the set of microscopic configurations compatible with a given value of the total energy, and canonical ensembles that corresponds to  microscopical configurations in which the total energy fluctuates around a given mean.
 A pivotal results of statistical mechanics is the equivalence of these ensembles in the thermodynamic limit, i.e., in the limit where the number of particle in the system is very large.
Similarly, in the theory of random graphs we distinguish between the $G(N,L)$ ensemble, which consists of all networks with $N$ nodes and a total of exactly $L$ links, and the $G(N,p)$ ensemble, which is formed by all networks of $N$ nodes and the total number of links being a Poisson distributed random variable with average $\langle L \rangle=p(N-1)$.
Exploiting the parallelism between statistical mechanics and theory of random graphs we can call the random graph ensemble  $G(N,L)$ a {\em microcanonical network ensembles} and the $G(N,p)$ graph ensemble a  {\em  canonical network ensemble}.
Similarly to statistical mechanics, the random graph ensembles $G(N,L)$ and $G(N,p)$ are,  in the thermodynamic limit  asymptotically equivalent as long as $L$ of the $G(N,L)$ ensemble and $p$ of the $G(N,p)$ ensemble are related by the equality $L=p(N-1)$.

It was shown in \cite{entropy1, entropy2, Kartik} that the parallel construction between  network ensembles can be extended to much more complex networks. In fact it is  possible to define {\em microcanonical network ensembles}  by imposing a set of hard constraints that must be satisfied by each  network in the ensemble  and  {\em canonical  network ensembles}, which satisfy soft constraints, i.e., the constraints are satisfied on average. The set of constraints might fix for example the degree sequence, the community structure or the spatial structure of networks embedded in space. 
   
A widely studied example of the {\em microcanonical network ensemble} is the configuration model \cite{MR} that fixes the degree sequence, i.e., degrees for all nodes in the networks. On the other hand, {\em canonical network ensembles} that impose soft contraints on the degree sequence  have been studied under different names (``hidden variable model" and ``fitness model") by the physics \cite{hv,Newman,Boguna} and statistics \cite{Snijders} communities.

In  a recent work \cite{Kartik}  it has been shown  that if the number of constraints is  extensive the microcanonical ensemble and its' conjugate canonical ensemble  are no longer equivalent. In particular, using a network  entropy measure, it was shown that a microcanonical ensemble has lower entropy than the conjugate canonical ensemble, even though the marginal probabilities take the same expression. An example of this difference was given by comparing the microcanonical ensemble of regular networks with fixed degree $k_i=c\in\mathbb{N}$ for all nodes $i=1,\ldots,N$ and the canonical Poisson network ensemble with average degree $\overline{k}_i=c$, for every $i=1,2,\ldots, N$, where the bar refers to the ensemble average.
It is easy to check that in this paradigmatic case, the entropy of the regular networks is smaller than the entropy of the Poisson networks with the same average degree.
The importance of such topological difference is also revealed by the observation that 
dynamical models defined on microcanonical network ensembles or corresponding canonical ones, display different critical  behavior.

The calculation for the entropy of arbitrary microcanonical ensembles was performed in \cite{entropy1, entropy2} using a Gaussian approximation and in   \cite{BianconiC, Annibale} by exact path integral approaches restricted to sparse networks and constraint degree sequence.
Here we show an extension of the exact results found in   \cite{BianconiC,Annibale} using the more transparent cavity method  \cite{Yedidia,Mezard} and derive the correspondence between the entropies of micro-canonical and conjugate canonical ensembles.

\section{Entropy of simple canonical network ensembles}

We first consider a canonical ensemble of {\it simple} networks, each  consisting of $N$ nodes and characterized by an adjacency matrix $\{\boldsymbol{a}\}\in\{0,1\}^{N\times N}$. A link between two nodes $i$ and $j$ may be present ($a_{ij}=1$) or absent ($a_{ij}=0$). The network is simple in that self-interactions are not permitted and that the adjacency matrix is symmetric.

Each network is described by its' probability distribution ${\cal P}(\{a\})=\prod_{i<j}\pi_{ij}({a_{ij}})$. The link between nodes $i$ and $j$ is present with probability $p_{ij}=\pi_{ij}(1)$ and is otherwise absent with probability $(1-p_{ij})=\pi_{ij}(0)$.

The ensemble is subject to  $\kappa=1\ldots M$ structural constraints, of the type
\begin{equation}
f_{\kappa}({\bf p})=F_{\kappa}\,,
\label{due}
\end{equation}
where $f_\kappa({\bf p})$ is a constraint function on the probability matrix $\{{\bf p}\}$, which consists of matrix elements $p_{ij}$, and $F_\kappa\in\mathbb{R}$ is the constraining value.

In accordance with the principle of maximal entropy \cite{Jaynes}, the link probabilities for this canonical ensemble are provided  by the maximization of  the Shannon  entropy of network ensembles \cite{Gfeller,Kartik},  
\begin{eqnarray}
\nonumber S[{\bf p}] &=& -\sum_{i<j}\sum_{\alpha=\{0,1\}}\pi_{ij}(\alpha)\ln(\pi_{ij}(\alpha))\\
\label{eq:con_ent}
&=& -\sum_{i<j} [p_{ij}\ln p_{ij}+(1-p_{ij})\ln(1-p_{ij})]\,,
\end{eqnarray}
subjected to the constraints of Eq. (\ref{due}). This optimization exercise gives rise to the maximal entropy canonical network ensemble, which is a generalization of the $G(N,p)$ random network ensemble \cite{Evolution1}.
The marginal probabilities $p_{ij}$ are given as the solution to the system of equations
\begin{equation}
\frac{\partial}{\partial p_{ij}} \left\{S[{\bf p}]+\sum_{\kappa=1}^{M} \lambda_{\kappa} f_{\kappa}({\bf p})\right\}=0\,,
\end{equation}
where the $\lambda_{\kappa}\in\mathbb{R}$ are Lagrange multipliers enforcing the constraints.

Let us consider the simple case of constraints on the expected degree of each node, i.e., we select $\overline{k}_i$, such that our $M=N$ constraints given by $(\ref{due})$ take the form
\begin{equation}
\sum_{j=1}^N p_{ij}=\overline{k}_i\,,\quad i=1,\ldots, N\,.
\label{l}
\end{equation}
The marginal probabilities $p_{ij}$ that satisfy Eq. $(\ref{l})$ are given as 
\begin{equation}
p_{ij}=\frac{e^{-\lambda_i-\lambda_j}}{1+e^{-\lambda_i-\lambda_j}}=\frac{\theta_i\theta_j}{1+\theta_i\theta_j}\,,
\end{equation}
with the Lagrange multipliers $\lambda_i$ fixed by Eq. $(\ref{l})$ and the variables $\theta_i=e^{-\lambda_i}$, which are commonly referred to as``hidden variables" \cite{hv,Newman,Boguna}.
In table \ref{table} we generalize this procedure to network ensemble satisfying a number of different structural constraints. 

\begin{table*}
\begin{ruledtabular}
\begin{tabular}{|c|c|c|}
 Constraints &Probabilities $p_{ij}/(1-p_{ij})$
 &\multicolumn{1}{c|}{Conditions} \\\hline 
$\begin{array}{c}\mbox{Given expected}\\ \mbox{ number of links }
 L\end{array}$ & $p/(1-p)$& $pN(N-1)/2=L$  \\\hline$\begin{array}{c}\mbox{Given expected}\\\mbox{ community structure} \{A_{q, q'}\}\end{array}$&
${W(q_i, q_j)}$ & 
$\begin{array}{c}\left. A(q, q')\right|_{q\neq
 q'}=\sum_{ij}p_{ij}\delta_{q_i, q}\delta_{q_j, q'}
 \\ A(q, q)=\sum_{i<j}p_{ij}\delta_{q_i, q}\delta_{q_j, q}\end{array}$
\\\hline $\begin{array}{c}\mbox{Given expected}\\\mbox{ degree
 sequence }\{\kappa_i\} \end{array}$&$\theta_i\theta_j$&$\kappa_i=\sum_{j}p_{ij}$ \\\hline
$\begin{array}{c}\mbox{Given expected}\\\mbox{ degree sequence } \{\kappa_i\}\\\mbox{ community structure } \{A(q, q')\}\end{array}$&${\theta_i
 \theta_j W(q_i, q_j)}$& $\begin{array}{c}\kappa_i=\sum_{j}p_{ij}\\
\left. A(q, q')\right|_{q\neq
 q'}=\sum_{ij}p_{ij}\delta_{q_i, q}\delta_{q_j, q'}
 \\A(q, q)=\sum_{i<j}p_{ij}\delta_{q_i, q}\delta_{q_j, q}\end{array}$ \\\hline $\begin{array}{c}\mbox{{\em Spatial networks}}\\\mbox{Given expected }\\ \mbox{degree
 sequence } \{\kappa_i\} \\ \mbox{and number of link at distance } d\in I_s, B_s\end{array}$& $\theta_i
\theta_jW(s_{ij}) $& $\begin{array}{c}\kappa_i=\sum_{j}p_{ij}\\B(s)=\sum_{ij}p_{ij}\chi_s(d_{ij})\end{array}$\\\hline
$\begin{array}{c}\mbox{Given expected}\\\mbox{ degree sequence }
 \{\kappa_i\} \\ \mbox{and number of triangles }\\\mbox{ for each node } \{T_i\} \end{array}$&
${\theta_i\theta_j e^{f_{ij}(\alpha_i+\alpha_j)+g_{ij}}}$ &
$\begin{array}{c} \kappa_i=\sum_{j}p_{ij}\\
 T_i=\sum_{jk}p_{ij}p_{jk}p_{ki}\\f_{ij}=\sum_{k}p_{ik}p_{kj}\\
 g_{ij}=\sum_kp_{ik}\alpha_kp_{kj} \end{array}$
 %\label{uno}
\end{tabular}
\end{ruledtabular}
\caption{Maximum-entropy networks ensembles with given set of
 constraints.The probability $p_{ij}$ of each link $(i,j)$ is given for network ensembles in which we imposed different types of constraints. This probabilities are expressed in terms of "hidden variables" of the
 ensembles $\{\theta_i\}$, 
 $W(q, q')$, $W(d)$, $\{\alpha_i\}$,  which are determined  by the 
 respective "conditions" specified in the table.   In the network ensembles with given community structure, the community of each node is associated with a Potts
 variable $q_i=1,\dots, Q={\cal O}(\sqrt{N})$. In the network ensemble embedded in a physical space the distance between the nodes  is binned in $L$ intervals $I_s\in[d_s,d_s+\Delta d_s)$ and it is indicated by
 a discrete variable $s_{ij}=s$ if the distance $d_{ij}$ between the nodes $i$ and $j$ satisfy $d_{ij}\in I_s$. The functions $\chi_s(d)$ are indicator functions of the intervals $I_s$, i.e. $\chi_s(d)=0,1$ and $\chi_s(d)=1$ if and only if $d\in I_s$.  
  }
\label{table}
\end{table*}

\section{Large deviations of canonical ensembles solved by the cavity method}

The constraints for canonical ensembles are satisfied only on average, it is therefore  relevant to investigate the probability of large fluctuations in these ensembles. The entropy for large deviations $\Omega[\{G_{\kappa}\}]$ of canonical ensembles is defined as
\begin{eqnarray}
\nonumber\Omega [\{G_{\kappa}\}] &=& -\lim_{N\rightarrow \infty}\frac{1}{N} \ln\Bigg[\sum_{\{a_{ij}\}}p_{ij}^{a_{ij}}(1-p_{ij})^{1-a_{ij}}\\
\label{omega_ent} &\times& \prod_{\kappa=1}^M\delta\Bigg(G_\kappa - g_{\kappa}({\bf a})\Bigg)\Bigg]\,,
\end{eqnarray}
where the delta function $\delta(\ldots)$ enforce the $\kappa=1,\ldots,M$ hard constraint
\begin{equation}
\label{eq:hq}
g_\kappa({\bf a}) = G_{\kappa}\,,
\end{equation}
with $g_\kappa({\bf a})$ being the constraining function specified on the adjacency matrix and $G_\kappa \in \mathbb{N}$ as the constrained value. The quantity $\Omega[\{G_{\kappa}\}]\geq0$ measures the probability that networks in a canonical ensembles   satisfy Eqs. $(\ref{eq:hq})$. If $\Omega[\{G_{\kappa}\}]$ is large, then this implies that the  probability that the networks in  the canonical ensemble satisfy  the topological constraints is large. Small values of $\Omega[\{G_{\kappa}\}]$, on the other hand, correspond to the large deviations of the canonical ensemble, i.e., there the networks satisfying the hard constraints are rare. 

The exact calculation of $\Omega[\{G_{\kappa}\}]$ has been performed using path integral methods \cite{BianconiC, Annibale} with linear hard constraints that fix the degree sequence.

Using the cavity method, we now demonstrate how to compute Eq. (\ref{omega_ent}) for more general cases of canonical ensemble and  hard constraints  fixing, for example, (i) the degree sequence, (ii) community structure, and (iii) number of links at a given distance.

\begin{figure}
\includegraphics[width=60mm, height=60mm]{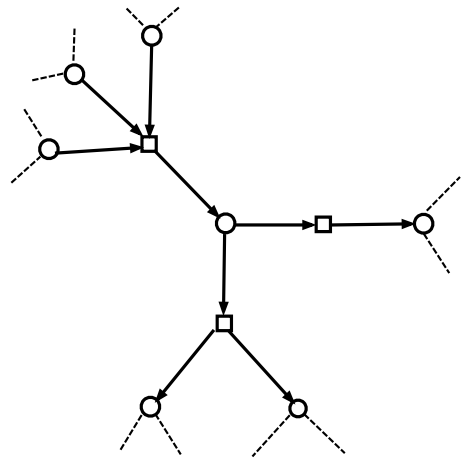}
\caption{Factor graph used for the cavity calculations. The variables nodes $\ell$ are indicated with circles and have a fixed connectivity $|\partial \ell|=3$ in the figure. The factor nodes, instead are indicated by squares. Their role is to  impose the hard constraints defined in Eqs.$(\ref{eq:hq})$. }
\label{factorg}
\end{figure}

In order to apply the cavity method to the calculation of $\Omega[\{G_{\kappa}\}]$ it is first necessary to describe the factor graph we will consider, which is depicted in Fig. $\ref{factorg}$. Following recent efforts to evaluate the number of loops in networks \cite{Semerjian1,Semerjian2,Valerie} we take the variables of the factor graph to be the matrix elements $a_{\ell}$ of the adjacency matrix, where index $\ell=1,\ldots,N(N-1)/2$ identifies each possible link of the undirected network \footnote{while extensions for directed networks are straightforward we shall focus, for the sake of clarity, on undirected networks only.}. The factor nodes, which are identified by Greek letters,$\alpha=1,2\ldots, M$ denote the $M$    topological constraints imposed on the network.
In particular, each factor node $\alpha$ is linked to a list of variables, which are identified by the set $\partial \alpha$. Likewise, variable $\ell$ is connected to a a set of  constraints, which is indicated as  $\partial \ell$. In our ensemble we assume that the number of constraints connected to a variable $\ell$ it is equal for each variable of the factor graph and given by $|\partial \ell|$.

The cavity method,  remains valid even for $M={\cal O}(N^2)$ but  the scaling $M={\cal O}(N)$ is necessary (as will become clear in the following derivation) to ensure that the entropy $\Omega[\{G_{\kappa}\}]$ remains finite.

\subsection{Large deviations of canonical ensembles with linear constraints}
The constraint given by Eq. $(\ref{eq:hq})$ now fixes the degrees of the factor nodes, i.e.,
\begin{equation}
K_{\alpha}=\sum_{\ell\in \partial \alpha}a_{\ell}\,,
\label{cs}
\end{equation}
with $\alpha=1,\ldots,M$ and factor node degree $K_\alpha\in\mathbb{N}$. Correspondingly, we can write Eq. (\ref{omega_ent}) as
\begin{eqnarray}
\nonumber \Omega[\{K_{\alpha}\}]&=&-\lim_{N\rightarrow \infty}\frac{1}{N} \ln\Bigg[\sum_{\{a_{\ell}\}}p_{\ell}^{a_{\ell}} (1-p_{\ell})^{1-a_{\ell}}\\&\times &\prod_{\alpha=1}^M\delta\left(K_{\alpha} - \sum_{\ell'\in\partial \alpha}a_{\ell'} \right)\Bigg],
\label{ld2}
\end{eqnarray}
Within the summation term on the first line of Eq. (\ref{ld2}) and for each value $a_\ell$, we introduce the unity identity $1=x^{a_\ell}\,x^{-a_\ell}$, which is parametrized by $x\geq0$. We can then define $\Omega_N[\{K_{\alpha}\},x]$ as
\begin{eqnarray}
\nonumber \Omega_N[\{K_{\alpha}\},x]&=&-\frac{1}{N} \ln\Bigg[\sum_{\{a_{\ell}\}}\left({p_{\ell}x}\right)^{a_{\ell}}({1-p_{\ell}})^{1-a_{\ell}}\\ &\times &\prod_{\alpha=1}^M\delta\left(K_{\alpha} - \sum_{\ell'\in\partial \alpha}a_{\ell'} \right)\Bigg]+\nonumber \\
&+&\frac{L}{N}\ln(x)
\label{Ox}
\end{eqnarray}
where $L$ is the total number of distinguishable links constraint by the constraint in Eq. $(\ref{cs})$. The introduction of the parameter $x$ at this stage is completely irrelevant and in fact the relation
\begin{equation}
\Omega[\{K_{\alpha}\}]=\lim_{N\to \infty}\Omega_N[\{K_{\alpha}\},x]
\label{limit}
\end{equation}
holds for all values of $x$. However, in what follows, we will focus on the particular limiting case where $x$ tends to $0$. Thus, we write that
\begin{equation}
\Omega[\{K_{\alpha}\}]=\lim_{x\to 0}\lim_{N\to \infty}\Omega_N[\{K_{\alpha}\},x]\,.
\end{equation}

The calculation of $\Omega[\{K_{\alpha}\},x]$ may be formulated in terms of the cavity method or the  Belief Propagation (BP) algorithm  \cite{Yedidia, Mezard}, aimed at determining $ \ln{\cal Z}$ with ${\cal Z}$ defined as in the following
\begin{eqnarray}
{\cal Z}&=&\sum_{\{a_{\ell}\}}\left({p_{\ell}x}\right)^{a_{\ell}}({1-p_{\ell}})^{1-a_{\ell}}\\ &\times &\prod_{\alpha=1}^M\delta\left(K_{\alpha} - \sum_{\ell'\in\partial \alpha}a_{\ell'} \right)\,,
\label{calZ}
\end{eqnarray}
where the entropy $\Omega_N[\{K_{\alpha}\},x]$ given by
\begin{equation}
N\Omega_N[\{K_{\alpha}\},x]=-\ln {\cal Z}+L\ln(x)\,.
\end{equation}
The  ``messages" of this BP algorithm are  sent between variable and factor nodes. In particular, we may define $\nu_{\ell\rightarrow \alpha}(a_{\ell})$ as the message sent from variable node $\ell$ to factor node $\alpha$ indicating the probability that matrix element $\ell$ takes value $a_{\ell}$, in absence of constraint $\alpha$. We correspondingly define $\hat\nu_{\alpha\rightarrow \ell}(a_{\ell})$ as the message that the  factor  node $\alpha$ sends to variable $\ell$ for the distribution of all variables connected to $\alpha$, except variable $\ell$. The BP update rules \cite{Yedidia, Mezard} take the form
\begin{widetext}
\begin{eqnarray}
\label{eq:cav_dist}\nu_{\ell\rightarrow \alpha}(a_{\ell})=\frac{1}{{\cal C}_{\ell,\alpha}}(p_{\ell}x)^{a_{\ell}}{(1-p_{\ell})}^{1-a_{\ell}}\prod_{\beta \in \partial \ell \setminus \alpha} \hat\nu_{\beta\rightarrow \ell}(a_{\ell}),\\
\label{nuh}
\hat\nu_{\alpha \rightarrow \ell}(a_{\ell})=\sum_{\{a_{\ell'}\}_{\ell'\in \partial \alpha \setminus \ell}}\delta\left(K_{\alpha}-a_\ell - \sum_{\ell'\in\partial\alpha\setminus \ell}a_{\ell'}\right)\prod_{\ell'\in \partial \alpha \setminus \ell}\nu_{\ell'\rightarrow \alpha}(a_{\ell'})\,,
\end{eqnarray}
\end{widetext}
where ${\cal C}_{\ell,\alpha}>0$ are a normalization constants.
To proceed, we make the ansatz that the cavity distribution satisfies a binomial form
\begin{equation}
\nu_{\ell\rightarrow \alpha}(a_{\ell})=h_{\ell,\alpha}^{a_{\ell}}(1-h_{\ell, \alpha})^{1-a_{\ell}}\,,
\end{equation}
which is parametrized by fields $h_{\ell, \alpha}\in\mathbb{R}$. Using integral representations of delta functions, we calculate the cavity messages given by $(\ref{nuh})$ as
\begin{equation}
\hat\nu_{\alpha\rightarrow \ell}(a_{\ell})=\int_{-\infty}^{\infty} \frac{dz}{2\pi} e^{-{\rm i}z\left[K_{\alpha}-a_{\ell}\right]}\prod_{\ell'\in \partial \alpha \setminus \ell}\Bigg[1-h_{\ell',\alpha}(1-e^{{\rm i}z})\Bigg]\,.
\end{equation}
Assuming self-consistently that the $h_{\ell' \alpha}$ are small, we approximate the product in the above equation as
\begin{eqnarray}
\nonumber \hat\nu_{\alpha \rightarrow \ell}(a_{\ell})&=&\int_{-\infty}^{\infty}\frac{dz}{2\pi} \exp\Bigg(-{\rm i}z\left[K_{\alpha}-a_{\ell}\right] \\
&-&\sum_{\ell'\in \partial \alpha \setminus \ell}h_{\ell',\alpha}(1-e^{{\rm i}z})\Bigg)\,,
\end{eqnarray}
which on suitable transformation of variables takes the form of Hankel's Contour Integral, giving
\begin{eqnarray}
\nonumber \hat\nu_{\alpha \rightarrow \ell}(a_{\ell})&=&\frac{1}{\Gamma(K_{\alpha}+1-a_{\ell})}\exp\left(h_{\ell,\alpha}-\sum_{\ell'\in \partial\alpha}h_{\ell',\alpha}\right)\\
&\times& \left(-h_{\ell, \alpha}+\sum_{\ell' \in \partial \alpha }h_{\ell', \alpha}\right)^{K_{\alpha}-a_{\ell}}\,.
\end{eqnarray}
Finally,  inserting the above result into Eq. (\ref{eq:cav_dist}), we get that $h_{\ell, \alpha}$ satisfied the recursion equation
\begin{equation}
h_{\ell,\alpha}=\frac{{p_{\ell}x}\prod_{\beta\in\partial\ell\setminus\alpha}\frac{K_{\beta}}{(-h_{\ell,\beta}+\sum_{\ell'\in\partial \alpha }h_{\ell',\beta})}}{1-p_{\ell}+{p_{\ell}x}\prod_{\beta\in\partial\ell\setminus\alpha}\frac{K_{\beta}}{(-h_{\ell,\beta}+\sum_{\ell'\in\partial \beta }h_{\ell',\beta})}}\,.
\label{bp0}
\end{equation}

Provided that every link exists with probability $p_{\ell} \neq 1$, we can choose the value of $x$ to be sufficiently small so as to approximate $h_{\ell,\alpha}$ by 
\begin{equation}
h_{\ell,\alpha}=\frac{p_{\ell}x}{1-p_{\ell}}\prod_{\beta\in\partial\ell\setminus\alpha}\frac{K_{\beta}}{(-h_{\ell,\beta}+\sum_{\ell'\in\partial \beta }h_{\ell',\beta})}\,.
\label{bp1}
\end{equation}
Since we have assumed  that every variable $\ell$ is linked to exactly $|\partial \ell|=2,3,\ldots$ factor nodes, the equation $(\ref{bp1})$ is solved for every value of $x \ll1 $
by the cavity field
\begin{equation}
h_{\ell,\alpha}={x}^{{1}/{|\partial \ell|}} \hat{h}_{\ell,\alpha}
\label{hs}
\end{equation}
where the cavity field $\hat{h}_{\ell,\alpha}$ satisfy the equation
\begin{equation}
\hat{h}_{\ell,\alpha}=\frac{p_{\ell}}{1-p_{\ell}}\prod_{\beta\in\partial\ell\setminus\alpha}\frac{K_{\beta}}{(-\hat{h}_{\ell,\beta}+\sum_{\ell'\in\partial \beta }\hat{h}_{\ell',\beta})}\,.
\label{bp2}
\end{equation}
Equations $(\ref{hs})$ and $(\ref{bp2})$ define  the cavity distributions $h_{\ell,\alpha}$ which are indeed small for sufficiently small value of $x$, as previously assumed.
Finally, using the BP algorithm \cite{Yedidia, Mezard} we can derive the  marginal distributions for the factor graph which are given by
\begin{widetext}
 \begin{eqnarray}
\nonumber P_{\ell}(a_{\ell}) &=& C_\ell^{-1}(p_{\ell}x)^{a_{\ell}}(1-p_{\ell})^{1-a_{\ell}}\prod_{\beta\in\partial\ell} \hat\nu_{\beta\rightarrow \ell}(a_{\ell}) \,,\\
\nonumber P_\alpha(\{a_{\ell^\prime}\}_{\ell^\prime\in\partial\alpha}) &=& C_\alpha^{-1}\delta\left(K_\alpha - \sum_{\ell^\prime\in\partial\alpha}a_{\ell^\prime}\right)\prod_{\ell^{\prime}\in \partial \alpha} \Bigg[ ({p_{\ell}x})^{a_{\ell}}{(1-p_{\ell})^{1-a_{\ell}}} \prod_{\beta\in\partial\ell^{\prime}\in \setminus \alpha}\hat\nu_{\beta\rightarrow\ell^\prime}(a_{\ell^\prime})\Bigg]\,,
\label{eq:marginal_dist}
\end{eqnarray}
\end{widetext}
where $C_\ell$ and $C_\alpha$ are normalization constants that satisfy
\begin{widetext}
\begin{eqnarray}
 C_{\ell}&=&\prod_{\alpha\in\partial\ell}\hat\nu_{\alpha\rightarrow\ell}(0) \Bigg[{p_{\ell}x}\prod_{\beta\in \partial \ell} \frac{K_{\beta}}{\sum_{\ell^\prime\in\partial\beta\setminus\alpha}h_{\ell^\prime,\beta}}+1-p_{\ell}\Bigg]\,,\\
C_{\alpha}&=&\pi_{(\sum_{\ell \in \partial \alpha}h_{\ell,\alpha})}(K_{\alpha}) \prod_{\ell^{\prime}\in \partial \alpha} (1-p_{\ell^{\prime}})\prod_{\beta\in \partial \ell^{\prime} \setminus \alpha}\hat{\nu}_{\beta\rightarrow \ell^{\prime}}(0)\,.\nonumber
\label{Cae}
\end{eqnarray}
\end{widetext}
The term $\pi_x(y)$ gives the probability for Poisson distributed random variable $y$ with average $x$. Following  \cite{Yedidia,Mezard}, in terms of our marginal distributions, the quantity $-\ln{\cal Z}$, with ${\cal Z}$ defined in Eq. $(\ref{calZ})$, may be expressed as the minimum of the Bethe free energy
\begin{widetext}
\begin{eqnarray}
 G_{{\rm Bethe}}[\{K_{\alpha}\}]=\sum_{\alpha=1}^M\sum_{\{a_{\ell}\}_{\ell\in \partial \alpha}}P_\alpha(\{a_{\ell}\})\ln\left(\frac{P_\alpha(\{a_{\ell}\})}{\psi_\alpha(\{a_{\ell}\})}\right) -\sum_{\ell=1}^{N(N-1)/2}(|\partial \ell|-1)\sum_{a_{\ell}=\{0,1\}}{P}_{\ell}(a_{\ell})\ln \left(\frac{{P}_\ell(a_{\ell})}{\phi_\ell(a_{\ell})}\right)\,,
 \label{G}
\end{eqnarray}
\end{widetext}
where $|\partial \ell|$ indicates the number of factor nodes connected to variable $\ell$ and
\begin{equation}
\psi_{\alpha} (\{a_{\ell}\})=\prod_{\ell \in \partial \alpha}({p_{\ell}x})^{a_{\ell}}({1-p_{\ell}})^{a_{\ell}}\,,
\end{equation}
and
\begin{equation}
\phi_\ell(a_{\ell})=(p_{\ell}x)^{a_{\ell}}(1-p_{\ell})^{1-a_{\ell}}\,.
\end{equation}
Inserting the expression for the marginal distributions, Eq. $(\ref{eq:marginal_dist})$, into Eq. $(\ref{G})$ we obtain the result \cite{Mezard,Semerjian1,Semerjian2} that
\begin{equation}
\label{calZ2}
-\ln {\cal Z}=-\sum_{\alpha=1}^{M}\ln C_{\alpha}+(|\partial \ell|-1)\sum_{\ell=1}^{N(N-1)/2} \ln C_{\ell}.
\end{equation}
Using the definition of entropy of large deviations Eq. $(\ref{Ox})$ and the expressions in Eq. $(\ref{Cae})$ for $C_{\alpha}$ and $C_{\ell}$, together with the Eqs. $(\ref{hs}-\ref{bp2})$ for the cavity fields, we get, for $x\ll 1$, that
\begin{widetext}
\begin{eqnarray}
\label{fi}
N\Omega_N[\{K_{\alpha}\},x]&=&-\sum_{\alpha=1}^{M}\ln C_{\alpha}+(|\partial \ell|-1)\sum_{\ell=1}^{N(N-1)/2} \ln C_{\ell}+L\ln(x)\,\nonumber \\
&=&-\sum_{\alpha=1}^{M}\ln \left[ \frac{1}{K_{\alpha}!}\left(x^{1/|\partial \ell| }\sum_{\ell \in \partial \alpha}\hat{h}_{\ell,\alpha}\right)^{K_{\alpha}}\exp\left(-x^{1/|\partial \ell| }\sum_{\ell\in \partial \alpha}\hat{h}_{\ell,\alpha}\right)\prod_{\ell \in\partial \alpha}(1-p_{\ell})\right]+\nonumber\\
&+&(|\partial \ell|-1)\sum_{\ell=1}^{N(N-1)/2} \ln \left[p_{\ell}\prod_{\beta\in \partial \ell}\frac{K_{\beta}}{\sum_{\ell'\in\partial \beta\setminus \alpha}\hat{h}_{\ell',\beta}}+1-p_{\ell}\right]+L\ln(x)\,.
\end{eqnarray}
\end{widetext}
Finally, going in the limit $x\to 0$ and $N\to \infty$ we get, according to Eq. $(\ref{limit})$
\begin{widetext}
\begin{eqnarray}
\label{fi2}
\Omega[\{K_{\alpha}\}]&=&\lim_{N\to \infty}\frac{1}{N}\left\{-\sum_{\alpha=1}^{M}\ln \left[ \frac{1}{K_{\alpha}!}\left(\sum_{\ell \in \partial \alpha}\hat{h}_{\ell,\alpha}\right)^{K_{\alpha}}\right]\right.+\nonumber\\
&+&\left.(|\partial \ell|-1)\sum_{\ell=1}^{N(N-1)/2} \ln \left[p_{\ell}\prod_{\beta\in \partial \ell}\frac{K_{\beta}}{\sum_{\ell'\in\partial \beta\setminus \alpha}\hat{h}_{\ell',\beta}}+1-p_{\ell}\right]-|\partial \ell|\sum_{\ell}\ln(1-p_{\ell})\right\},
\end{eqnarray}
\end{widetext}
where the cavity field $\hat{h}_{\ell,\alpha}$ are the solution of the BP Eq. $(\ref{bp2})$.

\subsection{Specific hard constraints}
We now consider a few specific cases for the hard constraints, which allow us to simply our expression Eq. $(\ref{fi2})$  further.

\subsubsection{Degree sequence}

Also known as the configuration model \cite{MR}, we consider constraints that fix the degree sequence, $(k_1,k_2,\ldots,k_N)\in\mathbb{N}^N$ for the network, where 
\begin{equation}
\label{conf_model}
k_i = \sum_{j=1}^N a_{ij}\,,
\end{equation}
with $i=1,\dots, N$. In terms of the factor graph, each factor node $\alpha$ fixes the degree for a specific node $i$ in the undirected network. Recalling that variable $\ell$ represents the tuple $(i,j)$ in the adjacency matrix, the variable is linked to $|\partial \ell|=2$ constraints that fix the degrees for nodes $i$ and $j$. Finally the cavity fields $h_{\ell,\alpha}$ can be written as  $h_{j,i}$, as we have identified factor node $\alpha$ with node index $i$ and, similarly, variable $\ell$ with nodes $i$ and $j$.

To simplify the expression Eq. $(\ref{fi2})$ for $\Omega[\{k_{i}\}]$ we introduce
\begin{equation}
\gamma_i=\sum_{j\neq i}^N \hat{h}_{j, i}\,.
\end{equation}
Using  Eq. $(\ref{bp2})$ it is easy to show that the variables  $\{\gamma_i\}$ satisfy the following equation 
\begin{equation}
\gamma_i=\sum_{j\neq i}^N\left(\frac{p_{ij}}{1-p_{ij}}\right)\frac{k_j}{\gamma_j-\hat{h}_{i, j}}\,,
\label{gamma}
\end{equation}
where $\hat{h}_{i,j}$ is given by the solution to Eq. $(\ref{bp2})$.
Finally, in the limit $x\to 0$, we get the exact result for the entropy of the large deviation of canonical network ensembles to be
\begin{eqnarray}
 \Omega[\{k_{i}\}]&=&\lim_{N\to \infty}\frac{1}{N}\left\{-\sum_{i=1}^N \ln\left[\frac{1}{k_i !}\gamma_i^{k_i}\right]+\right.\\
&+&\left.\sum_{\langle i,j \rangle} \ln \left[\frac{p_{ij}}{1-p_{ij}}\frac{k_ik_j}{\gamma_i\gamma_j}+ 1\right]-\sum_{\ell}\ln(1-p_{\ell})\right\}\,,\nonumber
\end{eqnarray}
where $\langle i, j \rangle$ indicates the sum over all links in the adjacency matrix.
If  $h_{j, i}\ll \gamma_i$ Eq. $(\ref{gamma})$ simplifies to
\begin{equation}
\gamma_i=\sum_{j\neq i}^N \frac{p_{\ell}}{1-p_{\ell}}\frac{k_j}{\gamma_j}\,,
\label{gamma2}
\end{equation}
which then gives in the diluted limit $p_{\ell}\ll 1$  the result \cite{BianconiC,Annibale} that
\begin{eqnarray}
\label{eq:ng} \Omega[\{k_{i}\}] &= &\lim_{N\to \infty}-\frac{1}{N}\left\{ \sum_{i=1}^N \ln\left[\frac{1}{k_i!}\gamma_i^{k_i}e^{-k_i}\right]\right\}\,,
\end{eqnarray}
for the configuration model.

\subsubsection{Community structure and degree sequence}
\label{commstr}
Suppose we assign to node $i$ a Pott's index $q_i=1,\dots, Q$  that indicates the community to which the node $i$ belongs.
In addition to the degree constraint given by Eq. (\ref{conf_model}), we also impose on the level of the adjacency matrix that
\begin{eqnarray}
\label{community_constraint}
A(q,q')&=&\sum_{i<j=1}^N(1-\frac{1}{2}\delta_{q,q'})\delta_{q,q_i}\,\delta_{q',q_j}\,a_{ij}\,, 
\end{eqnarray}
where $q<q'=1,\dots Q$. The total number of constraints is $M=N+Q(Q-1)/2$.

Each variable node $\ell$ in our factor graph is now linked to three factor nodes - two for constraining the degrees of nodes $i$ and $j$, separately, in the undirected network and a third one to enforce the community structure $q_i,q_j$. Similarly to the previous case we introduce
\begin{equation}
\label{eq:gamma_alpha}
\gamma_{\alpha}=\sum_{\ell\in \partial \alpha } \hat{h}_{\ell,\alpha}\,,
\end{equation}
where $\alpha\in\{i,j,(q_i,q_j)\}$ indicated the type of constraint. Given the cavity Eqs. $(\ref{bp2})$, it can be shown that the variables $\{\gamma_i\}$ satisfy the following equation
\begin{equation}
\gamma_{\alpha}=\sum_{\ell\in \partial \alpha}\frac{p_{\ell}}{1-p_{\ell}}\prod_{\beta \in\partial \ell\setminus \alpha} \frac{K_{\beta}}{\gamma_{\beta}-\hat{h}_{\ell,\beta}}\,,
\end{equation}
where $K_\beta \in \{k_i, k_j, A(q_i,q_j)\}$, depending on the value of $\alpha$ and the cavity fields $\hat{h}_{\ell,\alpha}$ satisfy the cavity Eqs. $(\ref{bp2})$. The entropy of large deviations $\Omega [\{K_{\alpha}\}]$ given by $(\ref{fi2})$ can be expressed  as 
\begin{eqnarray}
 \Omega[\{K_{\alpha}\}]&=&\lim_{N\to \infty}\frac{1}{N} \left\{-\sum_{\alpha=1}^M \ln[\pi_{\gamma_{\alpha}}(K_{\alpha})]\right.\nonumber \\
&+& \left.\sum_{i,j=1}^N \frac{p_{ij}}{1-p_{ij}}\left[\frac{k_i\,k_j\, A(q_i,q_j)}{\gamma_i\gamma_j\gamma_{(q_i, q_j)}}+1\right]\right.\nonumber \\
&-&\left.\sum_{\alpha} \gamma_{\alpha}-2\sum_{\ell}\ln(1-p_{\ell})\right\}.
\end{eqnarray}
In the case in which  $\hat{h}_{\ell,\beta}\ll \gamma_{\beta} $ and the network is diluted $p_{\ell}\ll 1$ we get
\begin{eqnarray}
 \Omega[\{K_{\alpha}\}]&=& \lim_{N\to \infty}\frac{1}{N}\left\{-\sum_{\alpha=1}^M \ln\left[\frac{1}{K_{\alpha}!}{\gamma_{\alpha}^{K_{\alpha}}}e^{-K_{\alpha}}\right]\right\}.
\end{eqnarray}
The value of $\Omega[\{K_{\alpha}\}]$ converges to a finite value in the limit of $N\rightarrow \infty$ only if the number of constraints $M$ is of the same order of magnitude as $N$, i.e., $M={\cal O}(N)$, in other words if the number of communities $Q={\cal O}(\sqrt{N})$. For $M={\cal O}(N^{\xi})$, with $\xi \in (1,2)$, we have that $\Omega[\{K_{\alpha}\}]\sim N^{\xi-1}$.

\subsubsection{Links at a given distance and degree sequence}
\label{links_dist}
Let us embed the $N$ nodes in a metric space, such that two nodes $i$ and $j$ are a distance $d_{ij}<D$ apart. We divide the interval $[0,D]$ into $L={\cal O}(N)$ intervals $I_s=[d_s, d_s+\Delta d_s)$ with $s=1,2,\ldots, L$ and $d_{s+1}=d_s+\Delta d_s$. The constraint for the number of links at a given distance is given by specifying a sequence of integers $B_1, B_2,\ldots B_L$ that satisfy
\begin{eqnarray}
B_s&=&\sum_{i<j}^N\chi_s(d_{ij})a_{ij}\,, 
\label{Bp1}
\end{eqnarray}
where $\chi_s(d_{ij})=1$ if  $d_{ij}\in I_s$ and $\chi_s(d_{ij})=0$ otherwise. The total number of constraints is in this case $M=N+L$.

Once again each variable $\ell$ is linked to $|\partial \ell|=3$ factor node constraints -- two for fixing the degrees  of node $i$ and $j$ and a third for the number of links $B_{s}$ in the interval $I_s$. We introduce the variables $\gamma_\alpha$ accoring to the definition
\begin{equation}
\label{eq:gamma_alpha}
\gamma_{\alpha}=\sum_{\ell\in \partial \alpha } \hat{h}_{\ell,\alpha}\,,
\end{equation}
with $\alpha\in\{i,j,s_{i,j}\}$.
These parameter satisfy the following equation
\begin{equation}
\gamma_{\alpha}=\sum_{\ell\in \partial \alpha}\frac{p_{\ell}}{1-p_{\ell}}\prod_{\beta \in\partial \ell\setminus \alpha} \frac{K_{\beta}}{\gamma_{\beta}-\hat{h}_{\ell,\beta}}\,.
\label{gammad}
\end{equation}
where the cavity field solve the cavity equation $(\ref{bp2})$. 
 The entropy of large deviations $\Omega [\{K_{\alpha}\}]$  given by $(\ref{fi2})$ can be expressed as 
\begin{eqnarray}
\nonumber  \Omega[\{K_{\alpha}\}]&=&\lim_{N\to \infty}\frac{1}{N}\left\{-\sum_{\alpha=1}^M \ln[\pi_{\gamma_{\alpha}}(K_{\alpha})]\right.\\
&+& \left.\sum_{i,j=1}^N\frac{p_{ij}}{1-p_{ij}}\left(\frac{k_i\,k_j\, B_{s_{i,j}}}{\gamma_i\,\gamma_j\,\gamma_{s_{i,j}}}-p_{ij}\right)\,\right.\nonumber \\
&-&\left.\sum_{\alpha} \gamma_{\alpha}-2\sum_{\ell}\log(1-p_{\ell})\right\}
\end{eqnarray}
where the subscript $s_{i,j}$ denotes the interval $s$ such that $d_{ij} \in I_s$.
Using $(\ref{gammad})$ in the limit of  spase networks with $p_{\ell}\ll1$ the entropy of large deviations simplifies and takes the form
\begin{eqnarray}
 \Omega[\{K_{\alpha}\}]&=&\lim_{N\to \infty}\frac{1}{N}\left\{-\sum_{\alpha=1}^M \ln\left[\frac{1}{K_{\alpha}!}{\gamma_{\alpha}}^{K_{\alpha}}e^{-K_{\alpha}}\right]\right\}
\label{d2}
\end{eqnarray}

The value of $\Omega[\{K_{\alpha}\}]$ converges to a finite limit for $N\rightarrow \infty$ only is the number of constraints $M$ is of the same order of magnitude as $N$, i.e. $M={\cal O}(N)$, i.e. only if $L={\cal O}(N)$
If $M\simeq N^{\xi}$ with $\xi\in(1,2)$ then $\Omega[\{K_{\alpha}\}]\sim N^{\xi-1}$.

\subsection{Special case for constraining degrees in sparse networks}
\label{sec:special}
Further simplifications for the expressions obtained in the previous section are possible when the constraining degrees $K_{\alpha}$ of sparse networks are the expected degrees over the canonical ensembles, i.e., $K_{\alpha}=\sum_{\ell\in\partial \alpha}p_{\ell}=k_{\alpha}$. The BP equations simplify to give 
\begin{equation}
h_{\ell,\alpha}=p_{\ell}\,.
\end{equation}
 Thus, Eq. (\ref{fi2}) reduces to
\begin{equation}
\Omega[\{k_{\alpha}\}]=-\lim_{N\to \infty}\frac{1}{N}\sum_{\alpha=1}^M\ln \pi_{k_{\alpha}}(k_{\alpha})\,.
\label{OLD}
\end{equation}
Since this is the minimum value of $\Omega$, we obtain that, for $M={\cal O}(N)$, the limit  $\lim_{N\to \infty}\Omega>0$ and therefore the canonical ensemble is not self-averaging in the thermodynamic limit.

\subsubsection{Degree sequence}
In the situation wherein only the degree sequence of the network is constrained, we have $K_i=\sum_{j=1}^Np_{ij}=k_i$, for all $i=1,\ldots,N$. The entropy $\Omega[\{k_i\}]$ of  the expected degrees  in the configuration model $\Omega[\{k_i\}]$ takes the form
 \begin{equation}
\Omega[\{k_i\}]=-\sum_{k>0}p_k\ln[\pi_k(k)]\,,
\end{equation}
where $p_k$ is the probability for a node to have degree $k$.

\subsubsection{Community structure and degree sequence}
As in Sec \ref{commstr}, each node $i$ is assigned a Pott's index $q_i=1,\dots, Q$ that indicates the community to which the node belongs, with $Q={\cal O}(\sqrt{N})$. The expected degree constraints take the form
\begin{eqnarray}
\label{eq:expdeg}k_i&=&=\sum_{j=1}^N p_{ij}\,,\\
A(q,q')&=&\sum_{i<j}^N(1-\frac{1}{2}\delta_{q,q'})\delta_{q,q_i}\,\delta_{q',q_j}\,p_{ij}\,,
\end{eqnarray}
for $i=1,\dots, N$ and $q<q'=1,\dots Q$
The total number of constraints is in this case $M=N+Q(Q-1)/2={\cal O}(N)$. The entropy $\Omega[\{k_i\},\{A(q,q')\}]$ takes the value
\begin{eqnarray}
 \Omega[\{k_i\},\{A(q,q')\}]&=&-\sum_{k>0}p_k\ln[\pi_k(k)]\\
&\hspace{-5mm}-& \hspace{-3mm}\lim_{N\to \infty}\frac{1}{N}\sum_{q< q'}^{Q} \ln[\pi_{A(q,q')}(A(q,q'))]\,.\nonumber
\end{eqnarray}

\subsubsection{Links at a given distance and degree sequence}
Following the setup of Sec. \ref{links_dist}, the constraints in terms of expected degrees are given by Eq. (\ref{eq:expdeg}) and
\begin{eqnarray}
B_s&=&\sum_{i<j}^N\chi_s(d_{ij})p_{ij}\,,
\label{Bp}
\end{eqnarray}
where $i=1,\dots,N$ and $s=1,\dots L$ and where $\chi_s(d_{ij})=1$ if  $d_{ij}\in I_s$ and $\chi_s(d_{ij})=0$ otherwise. We now express $\Omega[\{k_i\},\{B_p\}]$ as
\begin{equation}
\Omega[\{k_i\},\{B_p\}]= -\sum_{k>0}p_k\ln[\pi_k(k)]-\lim_{N\to \infty}\frac{1}{N}\sum_{s=1}^{L} \ln[\pi_{B_s}(B_s)]\,.
\end{equation}

\section{Entropy of simple microcanonical network ensembles } 

So far we have investigated the entropy of simple canonical network ensembles and large deviations therein. In this section we derive an expression for the entropy $\Sigma$ of a micro-canonical ensemble with linear constraints. Moreover, using the result of Eq. $(\ref{fi2})$ we relate $\Sigma$ to the entropy $\Omega$ of the most likely configuration of a  canonical ensemble when linear constraints are imposed.

Specifying $\kappa=1\,\ldots,M$ hard constraints on the adjacency matrix, as in Eq. $(\ref{eq:hq})$
, the micro-canonical ensembles' entropy $\Sigma$ is given by
\begin{equation}
\Sigma = \lim_{N\to \infty}\frac{1}{N}\ln Z_N\,,
\label{Sigma}
\end{equation}
where the partition function $Z_N$ is given by
\begin{equation}
Z_N=\sum_{\{a_{ij}\}}\prod_{\kappa=1}^M\delta\left(G_\kappa ({\bf a})- g_{\kappa}\right).
\label{ZN}
\end{equation}
In what follows we shall prove the following relationship, that
\begin{equation}
\Sigma= S^{\star}[{\cal P}]-\Omega^{\star}(\{G_{\kappa}\})\,,
\label{spo}
\end{equation}
where $S^{\star}[{\cal P}]$, given by Eq. (\ref{eq:con_ent}), is the Shannon entropy of the conjugated canonical ensemble. The term $\Omega^{\star}[\{G_{\kappa}\}]$ is the logarithm of the probability that a network in the conjugated canonical ensemble satisfies the hard constraints.

Physically, Eq. $(\ref{spo})$ implies that a network satisfying the hard constraints of Eq. $(\ref{eq:hq})$ belongs, with probability one, to the conjugated canonical ensemble. However, such networks make up only a fraction $e^{N\Omega^{\star}[\{G_{\kappa}\}]}$ of the total canonical ensemble measure.

\subsection{Proof of correspondence between canonical and micro-canonical entropies}
We now prove the relationship of Eq. $(\ref{spo})$, for the case of hard constraints specifying the degree sequence. 
In order to evaluate $(\ref{ZN})$ in this case we use the integral representation of the Dirac-delta functions, and we get
\begin{equation}
Z_N = \int \prod_{i=1}^N\frac{{\rm d}\omega_i}{2\pi} \exp\left(-\sum_{i=1}^N{\rm i}\omega_i k_i+\sum_{i<j}\ln[1+e^{{\rm i}\omega_i+{\rm i}\omega_j}]\right)\,,
\end{equation}
where with the change a variables $z_i=\omega_i-\omega_i^{\star}$
\begin{equation}
Z_N = \int \prod_{i=1}^N \frac{{\rm d}z_i}{2\pi} \,e^{F_N(\{z,\omega^\star\})}\,,
\end{equation}
with
\begin{eqnarray}
F_N(\{z,\omega^\star\})&=&-\sum_{i=1}^N\Bigg({\rm i}\omega_i^{\star}+{\rm i}z_i \Bigg)k_i + \sum_{i<j}\ln\Bigg[1+e^{{\rm i}\omega_i^{\star}+{\rm i}\omega_j^{\star}}\Bigg]\nonumber \\
\label{eq:F_z_ws}&+&\sum_{i<j}\ln\Bigg[1+p_{ij}\Big(e^{{\rm i}z_i + {\rm i}z_j}-1\Big)\Bigg]\,,
\end{eqnarray}
and the $\omega^\star$ variables chosen so as to satisfy the marginal probabilities for the canonical ensemble, i.e.,
 \begin{equation}
p_{ij}=\frac{e^{ {\rm i}\omega_i^{\star}+ {\rm i}\omega_j^{\star}}}{1+e^{ {\rm i}\omega_i^{\star}+ {\rm i}\omega_j^{\star}}}\,.
 \end{equation}
We observe that Eq. (\ref{eq:F_z_ws}) can be expressed 
 \begin{eqnarray}
\nonumber F_N(\{z,\omega^\star\})&=&S[{\cal P},\{\omega^\star\}]-\sum_i {\rm i}z_i\,k_i\\
 &+&\sum_{i<j}\ln(1+p_{ij}(e^{{\rm i}z_i+{\rm i}z_j}-1))\,.
 \end{eqnarray}
Therefore, with simple manipulations it can be shown that the partition function can be written as 
 \begin{eqnarray}
 Z_N&=&e^{NS^{\star}[{\cal P}]}\sum_{\{a_{ij}\}}p_{ij}^{a_{ij}}(1-p_{ij})^{1-a_{ij}}\prod_{i=1}^N\delta_{k_i,\sum_j a_{ij}}\nonumber \\
&=&\exp\Bigg(N[S^{\star}[{\cal P}]+\Omega_N[\{k_{i}\},1]\Big\}\Bigg)\,.
 \end{eqnarray}
Given the definition $(\ref{Sigma})$, this proves the relationship Eq. $(\ref{spo})$ between entropies of micro-canonical and conjugate canonical ensembles.
 
\subsection{Special cases for constraining degrees}
Following the simplification of Sec. \ref{sec:special} we assume that the constraining degrees $K_\alpha$ are expectation values of the canonical ensemble. Using Eq. $(\ref{spo})$ we get
 \begin{equation}
\Sigma= S^{\star}[{\cal P}]-\Omega^{\star}[\{k_{\alpha}\}]\,.
\label{gg}
 \end{equation}
 where $\Omega[\{k_{\alpha}\}]$ is given by $(\ref{fi2})$ where $k_{\alpha}$ are the expected degree of the canonical ensembles.
For sparse networks, we can use Eq. $(\ref{OLD})$ and $\Sigma$
 take the simple form
 \begin{equation}
\Sigma= S^{\star}[{\cal P}]+(|\partial \ell|-1)\sum_{k>0}{n}_{k}\ln\pi_k(k)\,.
\label{gg2}
 \end{equation}
 where $n_k$ is the probability that a random constraint enforce the degree $k$.
 
We note that when using a Gaussian approximation \cite{entropy1, entropy2} for network models with linear constraints, the value for the entropy  $\Sigma_G$ obtained for the micro-canonical ensembles is reasonably good, with an estimated error equal to
 \begin{eqnarray}
\nonumber |\Sigma-\Sigma_{G}| &=& \left|\frac{1}{N}\sum_{\alpha=1}^M \ln \frac{[k_{\alpha}e^{-1}]^{k_{\alpha}}\sqrt{2 \pi k_{\alpha}}}{k_{\alpha}!}\right|\\
&\leq &\frac{M}{N} |\ln{e^{-1}\sqrt{2\pi}}|=\frac{M}{N}0.08\,.
 \end{eqnarray}
 We conclude this section with the expressions for $\Sigma$ for a few specific constraints.
 \subsubsection{Degree sequence}
From  Eq. $(\ref{gg})$ we get in the case of the sparse configuration model 
\begin{equation}
\Sigma=S^\star[{\cal P}]+\sum_{k>0}p_k\ln[\pi_k(k)]\,,
\end{equation}
where $p_k$ is the probability of observing a node with degree $k$.

\subsubsection{Community structure and degree sequence}
In the ensemble with given degree sequence and a constraint on the number of links within and between communities $q=1,\ldots, Q$, for the total number of communities $Q={\cal O}(\sqrt{N})$. Here we obtain for sparse networks
\begin{eqnarray}
\nonumber \Sigma&=&S^\star[{\cal P}]+\sum_{k>0} p_k \log [\pi_k(k)]\\
&+& \lim_{N\to \infty}\frac{1}{N}\sum_{q,\leq q'=1}^Q\log [\pi_{A_{q,q'}}(A_{q,q'})]\,,
\end{eqnarray}
where $A_{q,q'}$ is given by Eq. (\ref{community_constraint}).

\subsubsection{Links at a given distance and degree sequence}
When the constraints are on the number of links at a specific distance and the degree sequence, the expression for the entropy of the micro-canonical ensemble takes the form 
\begin{equation}
\Sigma=S[{\cal P}]+\sum_{k>0} p_k \log [\pi_k(k)]+\lim_{N\to \infty}\frac{1}{N}\sum_{s=1}^L \log [\pi_{B_s}(B_s)]\,.
\end{equation}
where $B_s$ is given by Eq. $(\ref{Bp})$ valid for sparse networks.

\section{Conclusions}
In conclusion we have derived exact results for the large deviation  properties of canonical network ensembles and for the entropy of micro-canonical network ensembles in the case of simple networks with linear constraints. 

Our results apply to simple networks with given degree sequence, community structure and for networks embedded in a metric space. Our approach makes use of the transparent cavity method, which can also be extended to other types of constraints or directed networks.

Our calculations are valid even when the number of constraints scales like $M={\cal O}(N^2)$. Nevertheless, only in the case of a linear number of constraints, i.e., $M={\cal O}(N)$, can we ensure that the entropy  $\Omega[\{K_{\alpha}\}]$ remains finite in the limit $N\to\infty$.

Further inquiry will be directed toward the exact evaluation of the entropy of weighted networks and networks wherein the number of loops passing through each node is constrained.
Moreover, the relation between the information entropy of networks ensembles studied here and the von Neumann entropy, as introduced in \cite{BGS}, presents further scope for investigation.

 G. B.  acknowledge stimulating discussions with  A. Annibale and  A.C.C. Coolen.

\bibliographystyle{unsrt}

\end{document}